\documentclass[letterpaper]{article}
\usepackage{aaai25}
\usepackage{natbib}
\usepackage{times}
\usepackage{helvet}
\usepackage{courier}
\usepackage{lipsum}
\usepackage[hyphens]{url}
\usepackage{graphicx}
\usepackage{todonotes}
\usepackage{caption} 

\usepackage{enumitem}
\usepackage{amsmath}
\usepackage{adjustbox}
\usepackage{bibentry}

\usepackage{multirow}
\usepackage{booktabs}
\usepackage{amssymb}
\usepackage{svg}
\usepackage{algorithm}
\usepackage{algorithmic}
\usepackage{newfloat}
\usepackage{listings}

\floatstyle{ruled}
\newfloat{listing}{tb}{lst}{}
\floatname{listing}{Listing}

\title{Can LLMs Obfuscate Code? A Systematic Analysis of Large Language Models into Assembly Code Obfuscation}

\author {
    Seyedreza Mohseni\textsuperscript{\rm 1}\equalcontrib,
    Seyedali Mohammadi\textsuperscript{\rm 1}\equalcontrib,
    Deepa Tilwani\textsuperscript{\rm 2},
    Yash Saxena\textsuperscript{\rm 1}\thanks{Work done while remotely interning at KAI$^2$ Lab UMBC.},
    Gerald Ketu Ndawula\textsuperscript{\rm 1},
    Sriram Vema\textsuperscript{\rm 1},
    Edward Raff\textsuperscript{\rm 3},
    Manas Gaur\textsuperscript{\rm 1}
}
\affiliations {
    \textsuperscript{\rm 1}University of Maryland, Baltimore County, MD, USA\\
    \textsuperscript{\rm 2}University of South Carolina, SC, USA\\
    \textsuperscript{\rm 3}Booz Allen Hamilton, NY, USA\\
    \{mohseni1, m294, ysaxena1, geraldn1, sriramv1, edraff1, manas\}@umbc.edu, dtilwani@mailbox.sc.edu
}

\begin{document}
\maketitle

\begin{abstract}

Malware authors often employ code obfuscations to make their malware harder to detect. Existing tools for generating obfuscated code often require access to the original source code (e.g., C++ or Java), and adding new obfuscations is a non-trivial, labor-intensive process. In this study, we ask the following question: \textit{Can Large Language Models (LLMs) potentially generate a new obfuscated assembly code?} If so, this poses a risk to anti-virus engines and potentially increases the flexibility of attackers to create new obfuscation patterns. We answer this in the affirmative by developing the \textsc{MetamorphASM} benchmark comprising 
\textsc{MetamorphASM Dataset (MAD)} along with three code obfuscation techniques: dead code, register substitution, and control flow change.
The \textsc{MetamorphASM} systematically evaluates the ability of LLMs to generate and analyze obfuscated code using \textsc{MAD}, which contains
328,200 obfuscated assembly code samples. We release this dataset and analyze the success rate of various LLMs (e.g., GPT-3.5/4, GPT-4o-mini, Starcoder, CodeGemma, CodeLlama, CodeT5, and LLaMA 3.1) in generating obfuscated assembly code. The evaluation was performed using established information-theoretic metrics and manual human review to ensure correctness and provide the foundation for researchers to study and develop remediations to this risk. 
\end{abstract}
\begin{links}
\link{Code and Dataset}{https://github.com/mohammadi-ali/MetamorphASM}
\end{links}
\section{Introduction}
Writing metamorphic malware is non-trivial. It requires coding up multiple obfuscations, and it is a double-edged sword for malware delivery. Metamorphic malware is harder for Anti-Viruses to catch, but it also makes the malware larger because it needs to keep around the code for re-writing itself. Malware authors would generally prefer their programs to be smaller with less code so that they are not easily flagged in reporting or logging tools as something for Security Operations Center (SOC) analysts to investigate and thus discover the malware.

Large Language Models (LLMs) pose a new potential threat. Instead of including a large metamorphic code engine, they can simply call out over the internet to commercial LLMs to be rewritten one piece at a time. The code is far smaller since APIs for internet communication are built into and readily available on almost all operating systems.
We find it important to evaluate such possibilities given the wide array of attack patterns in malware and the exist of a live and motivated adversary~\cite{8553214,10.1145/3243734.3264418,10.1145/3473039,10.1145/2666652.2666666,7917369}.

A web call out to a Microsoft or Google domain is also innocuous and a strategy employed by sophisticated malware when possible to make their traffic nearly impossible for SOC analysts to detect. Being smaller and obfuscated, the malware can potentially reap the benefit of metamorphic engines without the cons. Access to API keys is a minor hindrance in this case, as theft of API keys for malicious use is a common attack pattern often achieved by simply scraping GitHub repositories \cite{lehmann2020everything}.

An important question, then, is how reliably can malware use these LLMs to re-write themselves. While a low error rate is tolerable as it imposes only minor opportunity cost to the attacker (they don’t successfully propagate to a device, but may have another chance to get to the same device later) and prior work has shown malware execution can be surprisingly robust to random code corruption~\cite{Fleshman2018}. We also wish to know if smaller local LLMs can perform this task, not because malware would use one (the size of the malware would explode and become obviously detectable in a log file), but it is important for security researchers to test locally what is achievable without being restricted to API calls that they may not have the budget for.

\begin{figure*}[t]
\centering
\includegraphics[width=0.8\textwidth]{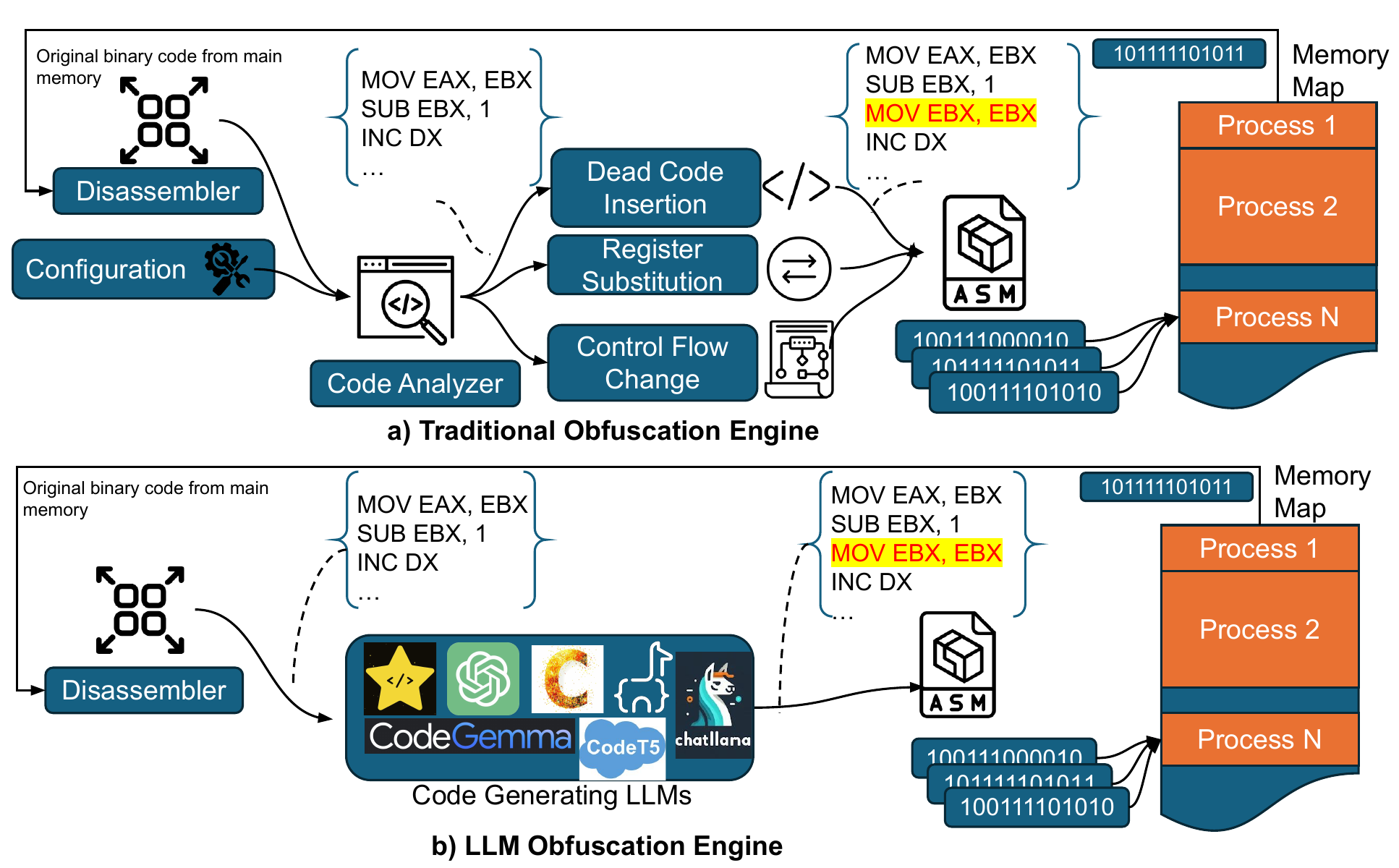} 
\caption{\label{fig:motivation_llm}\textbf{Traditional Obfuscation Engine v/s LLM Obfuscation Engine (\textsc{MetamorphASM Benchmark}).} 
\textbf{a)} The target code is fetched from the main memory disassembled, and fed to the code analyzer. The configuration unit provides metadata to the code analyzer. The code analyzer provides assembly code to obfuscation units, assembles them, and deploys binary into the main memory.
\textbf{b)} In the LLM obfuscation engine, we have to fetch from the main memory, disassemble it, and feed it into LLM. The LLM generates obfuscated code and sends it to the assembler to make a binary and deploy it to the main memory. 
 }
\end{figure*}

Figure~\ref{fig:motivation_llm} presents a schematic of the architecture of the classic obfuscation engine method compared to the proposed obfuscation engine using an LLM. In the classic obfuscation engine, the components can be classified as follows: (a) \textit{Assembler/Disassembler:} These components are responsible for converting binary code to assembly code or vice versa.
(b) \textit{Configuration Unit:} This unit provides necessary data to the code analyzer and obfuscator units to ensure efficient obfuscation. Typically, this unit is integrated within the mutation engine and supplies additional data to other units. (c) \textit{Code Analyzer:} This unit enhances the reliability of the obfuscation engine by providing extra information and analysis to the obfuscator units. (d) \textit{Obfuscator Units:} Depending on the purpose and complexity of the obfuscation engine, these components are responsible for obfuscating code based on code analysis. For more complex obfuscation, additional units are required.

We replace the \textit{code analyzer} and \textit{obfuscator units} with an LLM, as illustrated below in Figure~\ref{fig:motivation_llm}. The use of an LLM offers several potential advantages: (a) \textit{Ease of Generation and Training:} 
    LLMs require minimal development and debugging compared to the classic method, which involves extensive development and testing. (b) \textit{Platform Independence:} Unlike many classic obfuscation engines that are platform-dependent (e.g., Windows or Linux), LLMs are generally platform-independent and do not require special adjustments (c) \textit{Cost Efficiency:} 
    The cost of generating an LLM model is substantially lower than the development of a classic obfuscator engine with traditional programming languages such as C/C++ or JAVA. (d) \textit{Ease of Updating and Maintenance:} Updating and maintaining an LLM model is relatively straightforward.

\noindent Our contributions include the following:
\begin{itemize}[noitemsep]
    \item \textbf{The \textsc{MetamorphASM} dataset (MAD):} A dataset comprises 328,200 assembly code samples specifically crafted to test the ability of LLMs to perform code obfuscation. To our knowledge, this is the first assembly code obfuscation dataset, which provides researchers with a unique resource to perform a more detailed analysis of obfuscation strategies and evaluate the resilience of current detection technologies. 

    \item \textbf{Baseline Code-Generative Models:} We propose a series of baseline generative models, both a language model and LLMs, that are either trained, zero-shot prompted, or in-context-learned on our dataset, and evaluate them with the automatic scores and conduct a human review to inspect the obfuscation abilities. 
    \item We provide three distinct types of obfuscation each with 109400 samples (with an average code length from 399 to 507 for both original code and obfuscated/modified code): \textit{Dead Code Insertion, Register substitution, and Control Flow Change}, as the performance of LLMs can vary based on the specific obfuscation techniques used and the complexity of the code. 
       
\end{itemize}

\section{Background of Code Obfuscation Techniques}

Mathematically, we can show obfuscation with this \textbf{definition:} Code obfuscator is defined as a function $f$ that transforms original source code $P$ into an obfuscated version of source code $P'$.
Formally, this can be represented as
    $f : P \longrightarrow P'$.
Where $P$ is the space of all possible programs and $P'$ is the space of all possible obfuscated programs.

The obfuscation function $f$ must ensure that the obfuscated program $P'$ behaves identically to the original program $P$ for all inputs. So, we will have:
\begin{center}
  $\forall x \in \mathcal{X}~~  P(x) \simeq P'(x)$  
\end{center}
Where $\mathcal{X}$ represents the set of all possible inputs $x$ for which $P(x)$ and $P'(x)$ have a valid outcome without any error.

\begin{listing}[!h]
\caption{Original code}
\label{lst:1}%
\begin{lstlisting}
83C01C		ADD EAX, 28
8BE5	    MOV ESP, EBP
83E001		AND EAX, 1
0F94C1 		SETE CL
42		    INC EDX 
83EF01		SUB EDI, 1
56		    PUSH ESI
3BF9	    CMP EDI, ECX
57		    PUSH EDI
\end{lstlisting}
\end{listing}

\paragraph{Dead Code Insertion:}
In dead code insertion, malware inserts sections of code that are irrelevant to the program's normal operation. This technique can take various forms, such as adding redundant instructions, unused variables, or unreachable code branches. These additional code segments alter the structure and behavior of the malware without affecting its functionality, causing it to appear different each time it is executed. Consequently, traditional signature-based antivirus detection methods become less effective against metamorphic malware employing this technique. Listing-\ref{lst:1} shows a snippet assembly code which we called ``original code'' and Listing-\ref{lst:2} shows the original code after inserting dead code, such as ``NOP'' or ``MOV EDI, EDI''. Inserting dead code or garbage code, is one of the important method among the obfuscation techniques \cite{na2023dip}.

\begin{listing}[!h]
\caption{Original code after inserting dead code}
\label{lst:2}
\begin{lstlisting}
83C01C		ADD EAX, 28
90		    NOP           ;Dead code
8BE5	    MOV ESP, EBP
83E001		AND EAX, 1
8BFF	    MOV EDI, EDI  ;Dead code
0F94C1 		SETE CL
42		    INC EDX 
83EF01		SUB EDI, 1
90		    NOP           ;Dead code
56		    PUSH ESI
3BF9	    CMP EDI, ECX
57        PUSH EDI
\end{lstlisting}
\end{listing}

\paragraph{Register Substitution:}
In this technique, the malware replaces register names used within its instructions with alternative register names. For example, if the original malware code uses the ``EAX'' register for a specific computation, the register substitution technique might replace instances of ``EAX'' with ``EBX'' or another available register. Although the functionality of the code remains the same, altering the register names changes the code structure from its original form. Listing~\ref{lst:3} illustrates the register substitution technique \cite{balakrishnan2005code}.

\begin{listing}[!h]
\caption{Original code after register substitution}
\label{lst:3}
\begin{lstlisting}
83C31C		ADD EBX, 28  ;Swap EAX by EBX
8BE5	    MOV ESP, EBP
83E301		AND EBX, 1   ;Swap EAX by EBX  
0F94C1 		SETE CL
42		    INC EDX 
83EF01		SUB EDI, 1
56		    PUSH ESI
3BF9	    CMP EDI, ECX
57		    PUSH EDI
\end{lstlisting}
\end{listing}

\paragraph{Control flow change:}
In this technique, the malware rearranges the order of instructions in its code while preserving its original functionality. This rearrangement alters the sequence of instructions without changing the overall behavior of the malware. The purpose of instruction permutation is to disrupt the linear flow of the code and introduce variability in the instruction sequence (see Listing \ref{lst:4}, in comparison with Listing \ref{lst:1}. By constantly shuffling the order of the instructions, the malware presents a different code structure each time it is executed, making it challenging for antivirus programs to detect and analyze \cite{linn2003obfuscation}.

\begin{listing}[!h]
\caption{Original code after code flow change}
\label{lst:4}
\begin{lstlisting}
EB          JMP  sec1
sec3:
83EF01	    SUB EDI, 1
56	        PUSH ESI
3BF9        CMP EDI, ECX
57	        PUSH EDI
EB          JMP    sec4
sec2:
83E001	    AND EAX, 1
0F94C1 	    SETE CL
42	        INC EDX 
EB          JMP   sec3
sec1:
83C01C      ADD EAX, 28
8BE5        MOV ESP, EBP
EB          JMP  sec2 
sec4:	
\end{lstlisting}
\end{listing}

\section{\textsc{MetamorphASM Dataset (MAD)}}
The \textsc{MAD}, consisting of generated assembly code snippets, is generated through a four-step process:

\noindent \textbf{Step 1. The source of assembly codes.  } The source code comes from the extraction and disassembling of a large number of Dynamic Link Library and Program Executable, which Microsoft provided for Windows users, specifically Windows 7 and Windows 8.1. We used Windows Dynamic Link Library and Executive files because most of the metamorphic victims in the past and present are Windows users, and malware uses many standard libraries of .Net for reshaping code. We also use many open-source x64-based real assembly files or static libraries to generate assembly files.

\noindent \textbf{Step 2. Code extraction and pre-processing.} We use command prompt and open source software such as recompiling and disassembly tools to generate assembly code from original files. Most of the assembly code has a large section of data, which does not have a useful code for training LLMs. During the pre-processing step, we remove these sections and use only code sections to generate datasets. Another consideration for pre-processing is removing and purging near- and far-range JMPs or CALL instructions from the original code because all of these parameters are related to the local machines and temporary files, which causes a loss of generality concept during the training process of LLM. After cleaning up these large quantities of opcode, and after human evaluations and verification of large corpus, which took almost two months as full-time laboring, we break down this large corpus assembly code into small snippet assembly, each typically comprising twenty instructions. These snippets were then stored.

\noindent \textbf{Step 3. Obfuscating assembly codes.} After generating the clean assembly code snippets, the next step is to obfuscate each snippet using specific Python scripts. These scripts are designed to create three separate databases, each corresponding to a different obfuscation technique: dead code insertion, register substitution, and code flow alteration. To insert dead code, we use a dictionary of nearly 40 assembly instructions that do not affect the code's functionality but alter its structure. The script randomly inserts these ``neutral assembly instructions'' and saves the output in a Python dictionary as key-value pairs. For the code flow alteration dataset, the script reads each entry from the original database (created in step 2) and randomly rearranges parts of the code to obfuscate the original assembly snippet, then saves the result in the code flow change database. The final register substitution database involves reading the original database and randomly renaming specific registers (e.g., EAX, EBX, or ECX) by swapping them with other unused registers, with the results saved in the register dataset. By merging these three data into one unified  \textsc{MAD} dataset.

\noindent\textbf{Step 4. Final validations and verification.} The original code and the obfuscated code generated by three techniques is manually evaluated by human experts who have more than twenty years of experience in assembly and machine code development. 
In order to find any type of bug or defect, such as unwanted characters or wrong syntax, for removal. In the end, we package our datasets in Excel sheet format, which is ready to train models like CodeT5 and examine other powerful code-generating LLMs.

\subsection {Dataset Metrics}
The \textsc{MAD} focuses on three major obfuscation techniques: dead code insertion, register substitution, and control flow changes.
The \textsc{MAD} includes 109,400 entries for each obfuscation technique, structured as (original code, obfuscated code) pairs. The first item in each pair represents the original, unobfuscated code, while the second item contains the assembly code modified using one of the obfuscation techniques. Since \textsc{MAD} is designed for experiments with LLMs rather than real-world applications, each text entry (representing the original code) contains only twenty lines of assembly code. This size of code helps ensure minimal risk of future misconduct or misuse of the dataset.

For each dead code sample, we embed 4 to 5 dead code instructions into the original code and save it as its corresponding obfuscated code, resulting in obfuscated dead code that contains 24 to 25 lines of assembly code. For the register substitution sample, we apply register swapping to the original code and record it as its corresponding obfuscated code. Generally, each entry in the register substitution set will include at least one register swap, and the size of the swapped code remains the same as the original code. For control flow changes, we include 3 to 4 JMP instructions and their related labels, with the control flow of the program randomly altered for each entry. In all these obfuscated codes, the core functionality of the original code remains unchanged, but the structure of the code differs from the original.

\section{Models and Evaluation}

\paragraph{Models:} We conducted the benchmarking of \textsc{MetamorphASM} utilizing a diverse array of large language models (LLMs), which we categorized into open source (o) and proprietary (p) types, further divided into a mixture of experts (MoE) and non-mixture of experts (n-MoE) models. Access to proprietary LLMs was facilitated through APIs, which are computationally demanding and financially expensive. Therefore, we selected 15,000 assembly code samples - 5,000 per obfuscation mechanism - to ensure a fair and reasonable comparison - from our extensive repository of  300,000 examples. For our proprietary LLM evaluations, we included the GPT-4o-mini (p) \cite{achiam2023gpt}, an MoE model adept at handling complex assembly code patterns \cite{GPT-4o}. Competing against GPT-4o-mini, we employed the open-source DeepSeekCoder-v2 (o; 236B; MoE model)\cite{zhu2024deepseek} and Codestral~\cite{mistralCodestralHello} (o; 22B; n-MoE model), both of which have undergone extensive training on assembly codes and are proficient in generating such codes. Considering their established effectiveness in advanced coding tasks, CodeLLAMA (o; 34B) \cite{roziere2023code} and LLAMA 3.1\cite{metaIntroducingLlama} (p; 405B; n-MoE) were also included in our benchmarking process. Lastly, CodeGemma (o; 7B)\cite{team2024codegemma} and the trainable CodeT5 (o; 1.2B) \cite{wang2021codet5} were considered models suitable for conducting future white box studies into obfuscation using LLMs. Except for CodeT5, all the LLMs were subjected to examination using zero-shot prompting (refer to Table \ref{tab:zero-ct} for the zero-shot control flow change prompting template. The prompt templates for Dead Code and register substitutions can be found in the supplementary code materials.) and in-context learning, with test setups including 1, 3, 5, 10, and 15 shots (see Table \ref{tab:few-shot} for few shot prompting template).

\begin{table}[!ht]
\centering
\footnotesize
\begin{tabular}{|p{0.9\linewidth}|}
\hline
\multicolumn{1}{|c|}{\textbf{Zero Shot Control Flow Change Prompt}} \\ 
\hline
\textbf{Prompt:} Assembly Control Flow Change in obfuscation is a technique where the order of instructions is rearranged without altering the program's overall functionality. The goal is to make the code harder to understand and reverse-engineer. Control Flow Change leverages the fact that some instructions can be reordered safely if they are independent, meaning they do not depend on each other's results. Given the following original assembly code, determine which instructions can be safely reordered. Rearrange the identified independent instructions to achieve obfuscation. Just print the output code. \\
\textit{Original Assembly Code}: \\
\texttt{PUSH EDI} \\
\texttt{POP EDI} \\
\texttt{MOV EAX \dots} \\
\hline
\textbf{Response:} \\
\hline
\texttt{JMP sec1} \\
\texttt{sec4:} \\
\texttt{PUSH EAX \dots} \\
\hline
\end{tabular}
\caption{Zero Shot Control Flow Change Prompt Structure.}
\label{tab:zero-ct}
\end{table}

\begin{table}[!ht]
\centering
\footnotesize
\begin{tabular}{|p{0.9\linewidth}|}
\hline
\multicolumn{1}{|c|}{\textbf{Few Shot Prompt Structure}} \\ 
\hline
\textbf{Prompt:} Zero Shot Dead Code / Code Substitution / Control Flow Change + \textit{For Example:} \\
\textit{Original Code:} \\
\texttt{PUSH EDI} \\
\texttt{MOV EDI, DWORD PTR SS:[EBP+4]} \\
\texttt{PUSH 4 \dots} \\
\texttt{Augment k more examples for k few shot}\\

\hline
\textit{Original Assembly Code:} \\
\texttt{PUSH EDI} \\
\texttt{POP EDI} \\
\texttt{MOV EAX \dots} \\
\hline
\textbf{Response:} \\
\hline
\texttt{MOVZX EAX, AL} \\
\texttt{NEG EAX} \\
\texttt{ADD EAX, 0 \dots} \\
\hline
\end{tabular}
\caption{Few Shot Prompt Structure.}
\label{tab:few-shot}
\end{table}

\subsection{Evaluation}
To measure the obfuscation level, we consider two metrics. First, we compute the character-wise Delta Entropy ($\Delta$), which is a derived measure from Shannon Entropy. Analysts commonly use this measure as a first-pass analysis, with original code and obfuscated code. In the context of code obfuscation, it quantifies the complexity and diversity of the code. In fact, it gives us criteria regarding code mutation from original to obfuscated. For a given pair of assembly codes (original and obfuscated), we convert a snippet of original and generated assembly code into a sequence of symbols and apply entropy to these sequences. Then, we subtract the entropy of the original code from the generated code to measure the amount of obfuscation. It is defined as: 
\begin{center}
$\Delta H_{AB} = \frac{1}{N}\sum_{x\in n}^{N}(\mid H(A) - H(B)\mid)$
\end{center}
Where H(A) and H(B) stand for obfuscated code and original code. Second, we calculate the cosine similarity (CS) using its standard formulation to assess the similarity between the original and obfuscated code.

Functional correctness is an alternative method but it is impossible at this scale. Prior works use SMT solvers to validate compiler transforms are NP-Complete \citep{yang2024revisiting} and only guarantee soundness (i.e., no false equivalents) but not completeness (i.e., no missed equivalences), and tools used in practice expect some lifting from ASM to a higher representation \citep{furalabsGentleIntroduction}. For this reason, we use   Delta Entropy, which was proposed/evaluated by \citet{yang2024revisiting} 
, who found it an effective means of large-scale evaluation, as too dramatic a change in score is a strong indicator that the code has changed too significantly by either whole-re-writing (large change) or no edits (i.e., no change in score). There is a middle ground of plausibly valid transformations, and our use of manual expert evaluation over two months remediates this final uncertainty. 

\begin{table*}[!ht]
        \centering
        \adjustbox{max width=0.9\textwidth}{
        \begin{tabular}{lllllllllllllll}
        \hline
        & \multicolumn{2}{c}{0-Shot} & \multicolumn{2}{c}{1-Shot} & \multicolumn{2}{c}{3-Shot} & \multicolumn{2}{c}{5-Shot} & \multicolumn{2}{c}{10-Shot} & \multicolumn{2}{c}{15-Shot} \\
         \cmidrule(lr){2-3} \cmidrule(lr){4-5} \cmidrule(lr){6-7} \cmidrule(lr){8-9} \cmidrule(lr){10-11} \cmidrule(lr){12-13} 
         \textbf{LLMs} & \textbf{$\Delta (\%)$} & \textbf{CS} & \textbf{$ \Delta \%$} & \textbf{CS}& \textbf{$ \Delta \%$} & \textbf{CS}& \textbf{$ \Delta \%$} & \textbf{CS}& \textbf{$ \Delta \%$} & \textbf{CS}  & \textbf{$ \Delta \%$} & \textbf{CS}\\
        \hline
        GPT-4o-mini&26.90&0.93&21.00&0.95&17.50&0.95&20.70&0.95&19.30&0.96&22.33&0.95 \\
        GPT-3.5&10.22&0.93&17.34&0.90&3.82&0.80&0.98&0.83&5.74&0.77&2.88&0.77\\
        DeepSeekCoder-v2 & 19.50 & 0.99 & 22.00 & 0.99 & 25.50 & 0.99 & 26.40 & 0.99 & 27.00 & 0.99 & 27.50 & 0.99 \\
        Codestral & 30.25 & 0.95 & 15.47 & 0.96 & 14.53 & 0.97 & 13.67 & 0.97 & 12.10 & 0.98 & 11.93 & 0.98 \\ 
        Starcoder&61.35&0.68&45.55&0.97&56.25&0.97&56.40&0.97&58.70&0.97&57.28&0.97\\
        CodeGemma& 2.48& 0.30& 2.31& 0.31& 1.60& 0.38& 1.40& 0.40& 1.51& 0.40& 1.60& 0.4\\
        CodeLlama& 2.20  & 0.33 & 2.04& 0.32 & 1.57& 0.37& 1.39& 0.39 & 1.46 & 0.41 & 1.55 & 0.41\\
        LLama3.1&0.02&0.51&2.36&0.39&0.11&0.90&0.06&0.91&N/A&N/A&N/A&N/A \\
        \hline
        \multicolumn{13}{c}{Trained on \textsc{MAD}} \\
        \hline
        CodeT5	&0.06&	0.97&	-&-&	-&	-&-&	-&	-&	-&-&-\\

        \hline
        \end{tabular}
         }
        \caption{
            \textbf{Experimental results of the baseline models on the Dead Code Insertion obfuscation.} As we can observe, $\Delta$ entropy for Dead Code Insertion ranges from 10\% to 20\% due to inserting dead code into the original code for the top four models. The cosine similarity between 0.9 and 0.98 represents this technique's expected level of obfuscation. ``N/A'': LLM stopped generating assembly codes. ``-'': The model cannot be prompted with a few shot templates. 
        }
        \label{dead_results}
\end{table*}

\begin{table*}[!h]
        \centering
        \adjustbox{max width=\textwidth}{
        \begin{tabular}{lllllllllllllll}
        \hline
        & \multicolumn{2}{c}{0-Shot} & \multicolumn{2}{c}{1-Shot} & \multicolumn{2}{c}{3-Shot} & \multicolumn{2}{c}{5-Shot} & \multicolumn{2}{c}{10-Shot} & \multicolumn{2}{c}{15-Shot} \\
         \cmidrule(lr){2-3} \cmidrule(lr){4-5} \cmidrule(lr){6-7} \cmidrule(lr){8-9} \cmidrule(lr){10-11} \cmidrule(lr){12-13} 
         \textbf{LLMs} & \textbf{$ \Delta \%$} & \textbf{CS} & \textbf{$ \Delta \%$} & \textbf{CS}& \textbf{$ \Delta \%$} & \textbf{CS}& \textbf{$ \Delta \%$} & \textbf{CS}& \textbf{$ \Delta \%$} & \textbf{CS}  & \textbf{$ \Delta \%$} & \textbf{CS}\\
        \hline
        GPT-4o-mini & 26.30 & 0.93 & 1.20 & 0.92 & 0.46 & 0.92 & 5.08 & 0.92 & 10.21 & 0.92 & 13.86 & 0.91 \\
        GPT-3.5 & 2.30 & 0.94 & 0.96 & 0.90 & 2.07 & 0.92 & 0.24 & 0.91 & 0.15 & 0.90 & 0.60 & 0.89 \\
        DeepSeekCoder-v2 & 15.71 & 0.95 & 16.40 & 0.95 & 17.25 & 0.96 & 17.66 & 0.96 & 18.40 & 0.97 & 18.55 & 0.97 \\
        Codestral & 40.12 & 0.90 & 38.77 & 0.92 & 39.80 & 0.94 & 40.23 & 0.95 & 41.00 & 0.96 & 41.33 & 0.97 \\     
        Starcoder & 53.41 & 0.63 & 58.37 & 0.98 & 61.31 & 0.98 & 59.66 & 0.98 & 53.28 & 0.98 & 54.07 & 0.98 \\  
        CodeGemma  & 2.61  & 0.31 & 2.48& 0.36 & 2.57& 0.35 & 2.58 & 0.35  & 1.72 & 0.41 & 1.48 & 0.41  \\
        CodeLlama & 2.24  & 0.35 & 2.32 & 0.37& 2.34 & 0.37  & 2.39  & 0.36 & 1.67 & 0.40    & 1.47 & 0.40 \\
        LLama3.1 & 0.02 & 0.71 & 3.72 & 0.10 &2.40 & 0.26 & 0.09 & 0.56 & 0.01 & 0.54 &N/A&N/A\\
        \hline
        \multicolumn{13}{c}{Trained on \textsc{MAD}} \\
        \hline
        CodeT5 &	0.01&	0.99&	-&	-&	-&	-&	-&	-&	-&	-&-&-\\
       \hline
        \end{tabular}
         }
        \caption{ 
        \textbf{Experimental results of the baseline models on the Register substitution obfuscation.} As we can observe, $\Delta$ entropy for Register Substitution is very low in the top three models due to the only changing name of registers. The CS indicates that, in general, the similarity between two snippet codes is very high because of the swapping register names. 
        }
        \label{register_results}
\end{table*}

\begin{table*}[!h]
        \centering
        \adjustbox{max width=\textwidth}{
        \begin{tabular}{lllllllllllllll}
        \hline
        & \multicolumn{2}{c}{0-Shot} & \multicolumn{2}{c}{1-Shot} & \multicolumn{2}{c}{3-Shot} & \multicolumn{2}{c}{5-Shot} & \multicolumn{2}{c}{10-Shot} & \multicolumn{2}{c}{15-Shot} \\
         \cmidrule(lr){2-3} \cmidrule(lr){4-5} \cmidrule(lr){6-7} \cmidrule(lr){8-9} \cmidrule(lr){10-11} \cmidrule(lr){12-13} 
         \textbf{LLMs} & \textbf{$ \Delta \%$} & \textbf{CS} & \textbf{$ \Delta \%$} & \textbf{CS}& \textbf{$ \Delta \%$} & \textbf{CS}& \textbf{$ \Delta \%$} & \textbf{CS}& \textbf{$ \Delta \%$} & \textbf{CS}  & \textbf{$ \Delta \%$} & \textbf{CS}\\
        \hline
        GPT-4o-mini & 15.61 & 0.91 & 43.28 & 0.93 & 47.33 & 0.94 & 46.96 & 0.94 & 47.05 & 0.94 & 48.63 & 0.93 \\
        GPT-3.5 & 44.43 & 0.61 & 42.99 & 0.93 & 33.06 & 0.93 & 37.33 & 0.94 & 36.7 & 0.94 & 35.04 & 0.94 \\
        DeepSeekCoder-v2 & 49.20 & 0.99 & 50.40 & 0.99 & 51.35 & 0.99 & 51.50 & 0.99 & 52.00 & 0.99 & 52.55 & 0.99 \\
        Codestral & 45.77 & 0.60 & 47.23 & 0.82 & 50.12 & 0.90 & 52.84 & 0.93 & 55.30 & 0.95 & 54.91 & 0.95 \\ 
        Starcoder & 59.37 & 0.64 & 77.12 & 0.97 & 89.13 & 0.98 & 79.72 & 0.97 & 70.23 & 0.97 & 68.73 & 0.97 \\
        CodeGemma & 2.30& 0.32 & 1.82 & 0.24& 1.76 & 0.16& 1.78 & 0.16& 1.77 & 0.16 & 1.77& 0.16\\
        CodeLlama & 2.10& 0.35& 1.72& 0.28& 1.60 & 0.27 & 1.59& 0.23& 1.52 & 0.23  & 1.47 & 0.21\\
        LLama3.1 & 0.03 & 0.53 & 4.27 & 0.03 & 0.15 & 0.82 & 0.19 & 0.88 &N/A&N/A&N/A&N/A\\
        \hline
        \multicolumn{13}{c}{Trained on \textsc{MAD}} \\
        \hline
        CodeT5 &	0.12&	0.96&	-&	-&-&	-&	-&	-&	-&	-&-&-\\
        \hline
        \end{tabular}
         }
        \caption{\textbf{Experimental results of the baseline models on the Control Flow Change obfuscation.} As we can observe, $\Delta$ entropy for Control Flow Change is high in the top three models due to the insertion of the couple of JMPs instructions and section labels. Also, we have cosine similarity in the range 0.91 to 0.94, which shows 6\% to 9\% code obfuscation.     
        }
        \label{code_flow_results}
\end{table*}
\begin{table*}[!ht]
\centering
\resizebox{\textwidth}{!}{%
\begin{tabular}{lrrrrrrrr}
\toprule
Obfuscation & \textbf{GPT-4o-Mini} & \textbf{GPT-3.5} & \textbf{Starcoder} & \textbf{CodeLlama} & \textbf{CodeGemma} & \textbf{CodeT5} & \textbf{Codestral} & \textbf{DeepSeekCoder-v2} \\
\midrule
Deadcode & 1.67 & 3.00 & 6.00 & 7.67 & 8.67 & 7.33  & 4.67  & 1.33  \\
Register & 1.00  & 4.33  & 6.00 & 8.67  & 8.33  & 6.67  & 3.67    & 2.00  \\
Control Flow  & 1.67  & 4.33  & 7.00  & 8.33  & 8.33  & 5.67  & 3.67 & 1.33  \\
\bottomrule
\end{tabular}%
}
\caption{Human evaluation of various LLMs on the \textsc{MAD}. GPT-4o-mini and DeepSeekCoder-v2 were identified as the top two LLMs. There was ambiguity among evaluators about which model to rank as the third-best due to a tie between Codestral and GPT-3.5. Llama 3.1 was excluded from consideration due to its significantly low obfuscation rate and inability to generate obfuscated assembly code.}
\label{tab:human}
\end{table*}
\section{Results and Discussion}
\paragraph{Interpretation:} The \textsc{MAD} includes both original and obfuscated code, with an expected delta entropy range of around 10\%-20\%. This range is crucial for defining an effective obfuscation engine; a delta entropy exceeding this range risks altering the code's functionality, while a value below 10 percent indicates minimal obfuscation. The range was defined after three human experts examined the code obfuscation from eight LLMs and picked GPT-4o-mini as the closest to human-performed code obfuscation (see Table \ref{tab:human}). 
Additionally, maintaining a cosine similarity above 0.9 is essential, as it confirms the preservation of functional similarity between the original and obfuscated code, thereby serving as a measure of the obfuscation's success in maintaining the code's integrity without compromising its functionality. The threshold for cosine similarity was set following human evaluation, where experts reviewed the top three LLMs across three obfuscation techniques. We calculated the cosine similarity between the original and obfuscated code produced by top-3 LLMs, achieving an average of 0.9.

\paragraph{Discussion:} In a comparative analysis of general-purpose LLMs, LLAMA 3.1 exhibits notably underperformed, especially in control flow change techniques, where it only achieves a 4.27\% entropy rate in single-shot scenarios, highlighting its inadequate code mutation capabilities. In more complex tasks requiring 10 to 15 shots, LLAMA 3.1 fails to generate any valid assembly instructions and demonstrates considerable variability in cosine similarity, deviating from the expected range of 0.90 to 0.97. In contrast, GPT-4o-mini demonstrates robust performance across both entropy and cosine similarity metrics, excelling particularly in control flow change obfuscation with high entropy due to the insertion of numerous JMP and Section commands. Following closely, dead code insertion shows commendable results, and register substitution ranks third, indicating lower entropy typically associated with changes in one or two register names. Although GPT-3.5 outperforms LLAMA 3.1, it slightly trails GPT-4o-mini but maintains a cosine similarity within the desired range of 0.90 to 0.97.

Codestral stands out among specialized large language models for its effective performance in dead code and control flow change tasks, with cosine similarity values ranging from 0.90 to 0.98 (see Tables \ref{dead_results}, \ref{code_flow_results}). However, it struggles with register substitution, indicating difficulties in modifying register names effectively Table \ref{register_results}. DeepSeekCoder follows with higher performance, reflected in its elevated cosine similarity scores. These suggest that it accurately replicates the original assembly code, hinting at its specialized training in assembly language, making it a proficient obfuscator across all techniques. In contrast, CodeGemma and CodeLLAMA show inadequate results in the three obfuscation techniques, primarily due to their training in high-level programming languages like C/C++/C\#, Java, Rust, and Python rather than assembly. This leads to significant inaccuracies and irrelevant outputs. StarCoder, while capable of generating assembly code, demonstrates a high variability in entropy, suggesting it understands assembly but fails to consistently obfuscate at this level. Similarly, control flow change is the most effective obfuscation technique across specialized LLMs, followed by dead code insertion. Register substitution appears weaker, with a higher susceptibility to de-obfuscation. CodeT5, despite being fine-tuned, produces high cosine similarity in register substitution, around 0.98, indicating a strong resemblance to the original code. Yet, its low entropy suggests minimal to no actual obfuscation, often merely reproducing the original code as obfuscated.
\textbf{Human Evaluation:}
We designed three criteria to assess the effectiveness of obfuscation techniques in various code-generating language models, each rated on a scale from 1 to 8. The criteria are: (a) ranking the eight outputs based on the insertion of ineffective code, (b) ranking based on the substitution of registers, and (c) ranking based on the rearrangement of code sequences. We chose 200 random assembly code samples for this evaluation, conducted by three experts specializing in malware analysis. The results are in Table \ref{tab:human}, where a lower score signifies higher-quality obfuscation.

\section{Related Work}
Numerous classical software obfuscation techniques have been developed to safeguard against software tampering and reverse engineering, thereby preventing unauthorized access to source codes \cite{nagra2009surreptitious,hosseinzadeh2018diversification,xu2020layered,ahire2020mechanisms}. Tools such as LOKI and OBFUS reflect practical applications of these methodologies \cite{schloegel2022loki,kang2021obfus}. The LLVM (Low-Level Virtual Machine) is particularly notable for its flexibility and extensibility in both obfuscation and de-obfuscation, commonly employing techniques like control flow alteration and dead code insertion \cite{junod2015obfuscator,garba2019saturn}. This study extends existing research by examining the potential of LLMs to develop obfuscation engines \cite{gupta2023chatgpt}. While much of the existing research in this field concentrates on detection and defense, this effort focuses on utilizing LLMs trained in high-level programming languages, which are traditionally easier for experts to manage and understand \cite{muennighoff2023octopack}. However, training these models presents significant challenges due to the syntactic diversity and complexity of programming paradigms, requiring substantial resources \cite{hou2023large}. Our dataset and trained models can test the robustness and reliability of traditional and LLM-based detection systems. \textsc{MAD} enables studying other challenges in malware dataset construction, such as lack of diversity, data augmentation, and availability~\cite{saul2024is,liu2024assemblage,MALDICT,JOYCE2023102921,TirthCAMLIS} --- but are beyond the scope of this article. 

Our method has leveraged a heuristic approach to large-scale evaluation of code/malware. While provably correct equivalence is preferable,  it is not tenable at this scale of work. Prior work has considered the modifying the raw assembly of a program at high computational cost by leveraging domain knowledge re-writers to maintain semantic preserving changes, but note that this may not correctly handle any self-referential code (e.g., a checksum used to branch)~\cite{10.1007/978-3-031-43412-9_16}. This cost increases in our setting of LLM based code changes as we can not leverage any domain expert system to accelerate an equivalency check~\cite{8661180}. Since general code equivalence is NP-hard, such domain knowledge is required and known to be sound, but not complete~\cite{10.1145/1993498.1993533}. 

\section{Conclusion}
In this work, we provide \textsc{MAD}, which is a dataset for assembly code obfuscation for prompting and in-context learning of the LLM. Our dataset can obfuscate snippets assembly code by applying three major obfuscation techniques: a) inserting dead instructions code, b) register substitution, and c) changing control flow. For the purpose of demonstrating the trainability and reliability of our dataset, we tested our dataset by pre-training and prompting a couple of well-known models, such as the GPT family, CodeLLAMA, CodeGemma, Starcoder, Codestral, and DeepSeekCoder-v2. We also fine-tuned CodeT5 on our dataset, leveraging its open-source nature and transparent, white-box architecture. In order to measure the performance of models, we used Cosine similarity and Shannon entropy to measure the level of obfuscation between the original code and the generated code by the models. As shown in this paper, surprisingly, the GPT family (which is not a special coder LLM) has outstanding performance for obfuscation assembly code over even specialized coder LLM such as DeepSeekCoder-v2, Codestral, CodeLLAMA, CodeGemma, and Starcoder. The experiments demonstrated that even the pre-trained models show high performance on the obfuscation task, but it does not necessarily lead to high grounding performance, and GPT is still dominant.

\section*{Acknowledgements}
We acknowledge the support from UMBC
Cybersecurity Leadership – Exploratory Grant Program. Any opinions, conclusions, or recommendations expressed in this material are those of the authors and do not necessarily reflect the views of UMBC or Booz Allen Hamilton.

\bibliography{aaai25}

\end{document}